\documentclass[aps,prx,10pt,twocolumn]{revtex4-2}
\usepackage[ascii]{inputenc}
\usepackage{amsmath,amssymb,amsfonts,amsthm}
\usepackage[braket,qm]{qcircuit}
\usepackage{tikz}
\usetikzlibrary{matrix,chains,positioning,decorations.pathreplacing,arrows,arrows.meta,fit,patterns}
\usepackage{braket}
\usepackage[scientific-notation=true]{siunitx}
\usepackage{graphicx}
\usepackage[caption=false]{subfig}
\usepackage{hyperref}
\hypersetup{bookmarks=false,colorlinks=true,linkcolor=blue,
citecolor=blue,filecolor=black,urlcolor=blue}

\newcommand{\ud}{\mathrm{d}}

\newcommand{\R}{\mathbb{R}}
\newcommand{\C}{\mathbb{C}}

\newcommand{\vect}[1]{\boldsymbol{#1}}

\DeclareMathOperator{\e}{e}

\graphicspath{{figures/}}

\begin{document}

\title{Solving Differential Equations via Continuous-Variable Quantum Computers}

\author{Martin Knudsen}
\author{Christian B.~Mendl}
\affiliation{Technical University of Munich, Department of Informatics, Boltzmannstra{\ss}e 3, 85748 Garching, Germany}

\date{\today}

\begin{abstract}
We explore how a continuous-variable (CV) quantum computer could solve a classic differential equation, making use of its innate capability to represent real numbers in qumodes. Specifically, we construct variational CV quantum circuits [Killoran et al., Phys.~Rev.~Research 1, 033063 (2019)] to approximate the solution of one-dimensional ordinary differential equations (ODEs), with input encoding based on displacement gates and output via measurement averages. Our simulations and parameter optimization using the PennyLane / Strawberry Fields framework demonstrate good convergence for both linear and non-linear ODEs.
\end{abstract}

\maketitle

\section{Introduction}
\label{sec:Introduction}

Continuous-variable (CV) quantum computing \cite{Lloyd_1999, Braunstein2005, Weedbrook2012} uses bosonic modes (``qumodes'') as constitutive building blocks, analogous to qubits of a digital quantum computer. However, instead of a discrete basis set, in the CV formalism a position operator eigenstate $\ket{x}$ for example is described by a real number $x \in \R$. Thus CV quantum computing becomes a promising approach for computational tasks involving real-valued quantities. In this study, we propose and investigate a CV method for solving differential equations, which are ubiquitous in scientific and engineering fields and have recently attracted interest as potential application domain of quantum computers \cite{Xin2018, Arrazola_2019, Lloyd2020, Liu2020, Zanger2020}. Here we utilize a variational circuit approach \cite{Killoran2019}, which is very flexible due to its structural similarity with artificial neural networks and overcomes the apparent dichotomy between \emph{nonlinear} differential equations and linear quantum operations. An additional motivation for the analog paradigm are theoretical arguments that certain physical systems are inherently insusceptible to digital simulation \cite{Boche1}.

For concreteness, we aim to solve a one-dimensional initial value problem (IVP) represented as
\begin{equation}
y'(x) = f(x, y(x)), \quad y(x_0) = y_0
\label{eq:generalODE}
\end{equation}
with $x \in \Omega \subseteq \R$, a given real-valued right side $f: \Omega \times \R \to \R$ and initial condition at $x_0$, and $y: \Omega \to \R$ the sought solution. Our general strategy is to set up a variational quantum circuit which will approximate $y(x)$ after optimizing its parameters, see Sect.~\ref{sec:neural_network_variational_circuit}. The topology is inspired by classical artificial networks, as explained in the following Sect.~\ref{sec:classic_artificial_neural_network}.

\section{Classic artificial neural network ansatz}
\label{sec:classic_artificial_neural_network}

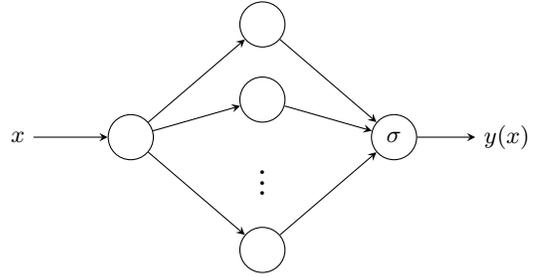
\begin{figure}[!ht]
\centering
\begin{tikzpicture}[>=stealth, neuron/.style={draw, circle, minimum size=0.6cm, inner sep=0}]
\node         (x)   at (-1.5, 0)    {$x$};
\node[neuron] (n0)  at (0, 0)       {};
\node[neuron] (n1a) at (1.75,  1.5) {};
\node[neuron] (n1b) at (1.75,  0.5) {};
\node         (d)   at (1.75, -0.5) {\Large $\vdots$};
\node[neuron] (n1c) at (1.75, -1.5) {};
\node[neuron] (n2)  at (3.5, 0)     {$\sigma$};
\node         (y)   at (5, 0)       {$y(x)$};
\draw[->] (x)   -- (n0);
\draw[->] (n0)  -- (n1a);
\draw[->] (n0)  -- (n1b);
\draw[->] (n0)  -- (n1c);
\draw[->] (n1a) -- (n2);
\draw[->] (n1b) -- (n2);
\draw[->] (n1c) -- (n2);
\draw[->] (n2)  -- (y);
\end{tikzpicture}
\caption{Classic neural network approach for solving an ordinary differential equation. Adapted from \cite{ODEClassic}.}
\label{fig:classicODENN}
\end{figure}

As blueprint for our quantum ansatz, we first explain how a classic artificial neural network (ANN) \cite{LeCun2015} can be  optimized to represent the solution of an ordinary differential equation (ODE) \cite{ODEClassic}, see Fig.~\ref{fig:classicODENN}. In mathematical terms, a traditional ANN consists of a composition of layers, $\mathcal{L}_L \circ \cdots \circ \mathcal{L}_2 \circ \mathcal{L}_1$, where a single layer implements an affine transformation followed by a pointwise nonlinear ``activation function'' $\sigma$:
\begin{equation}
\mathcal{L}_\ell: \R^{n_\ell} \to \R^{n_{\ell+1}}, \quad x \mapsto \sigma(W_\ell \cdot x + b_\ell)
\label{eq:ANN}
\end{equation}
for $\ell \in \{1, \dots, L\}$, $W_\ell \in \R^{n_{\ell+1} \times n_\ell}$ and $b_\ell \in \R^{n_{\ell+1}}$. The so-called weight matrices $W_\ell$ and bias vectors $b_\ell$ are the adjustable (``trainable'') parameters of the network. The gradient of some target function (depending on the network output) with respect to these parameters can be efficiently computed via backpropagation.

Solving \eqref{eq:generalODE} can now be reformulated as minimization of a cost function, denoted $C_\text{ODE}$, by discretizing the domain $\Omega$ by a set of points $\{ x_i \}_{i=1,\dots,N}$, and defining (cf.~\cite{ODEClassic})
\begin{equation}
C_\text{ODE} = \big(y(x_{0}) - y_{0}\big)^2 + \sum_{i=1}^N  \big(y'(x_i) - f(x_i, y(x_i))\big)^2.
\label{eq:ODELossFunction}
\end{equation}
Note that the derivative $y'(x)$ appears inside the cost function, which is a particular feature when considering an ODE. Consequently, during backpropagation the simultaneous partial derivatives of the output with respect to the input $x$ \emph{and} a network parameter $\theta$ are needed:
\begin{multline}
\frac{\partial C_\text{ODE}}{\partial \theta} =  2\big(y(x_{0}) - y_{0}\big) \frac{\partial y(x_{0})}{\partial \theta} \\ + \sum_{i=1}^N  2\big(y'(x_i) - f(x_i, y(x_i))\big) \\ \times \left(\frac{\partial^2 y(x_i)}{\partial \theta \partial x} - \frac{\partial f(x_i, y(x_i))}{\partial y(x_i))} \frac{\partial y(x_i))}{\partial \theta} \right).
\label{eq:partialDerivativeBackpropagation}
\end{multline}

In this work, the $\{ x_i \}$ were chosen to be $20$ evenly spaced points within the domain $\Omega$.

\section{Neural network-inspired variational quantum circuit}
\label{sec:neural_network_variational_circuit}

We make use of quantum gates available in the CV quantum computing formalism, namely displacement gates $D(\alpha)$, rotation gates $R(\phi)$, squeeze gates $S(r)$, beamsplitters $\mathrm{BS}(\theta)$ and Kerr gates $K(\kappa)$; appendix~\ref{sec:CV_gates} summarizes their mathematical definitions.

\begin{figure}[!ht]
\centering
\Qcircuit @C=0.4em @R=0.7em {
\pureghost{0} & \pureghost{0} & \pureghost{0} & \ustick{\text{QNN layer}} & \pureghost{0} & \pureghost{0} & \pureghost{0}\\
&\gate{D(\alpha_1)} &\multigate{3}{U_1(\vect{\phi}_1,\vect{\theta}_1)} & \gate{S(r_1)} & \multigate{3}{U_2(\vect{\phi}_2,\vect{\theta}_2)} & \gate{K(\kappa_1)} &\qw \\
&\gate{D(\alpha_2)} &\ghost{U(\vect{\phi}_1,\vect{\theta}_1)} & \gate{S(r_2)} & \ghost{U(\vect{\phi}_2,\vect{\theta}_2)} & \gate{K(\kappa_2)}&\qw \\
\pureghost{.} & \raisebox{.3em}{\vdots} &\pureghost{U(\vect{\phi}_1,\vect{\theta}_1)} & \raisebox{.3em}{\vdots} & \pureghost{U(\vect{\phi}_2,\vect{\theta}_2)} &  \raisebox{.3em}{\vdots} &\pureghost{0}\\
&\gate{D(\alpha_n)} &\ghost{U(\vect{\phi}_1,\vect{\theta}_1)} & \gate{S(r_n)} & \ghost{U(\vect{\phi}_3,\vect{\theta}_3)} & \gate{K(\kappa_n)} & \qw 
\gategroup{2}{2}{3}{6}{.7em}{^\}}
}
\caption{Variational circuit for implementing a single neural network layer on a CV quantum computer, consisting of two universal interferometers as well as one displacement, squeeze and Kerr gate for each wire. Adapted from \cite{Killoran2019}.}
\label{fig:QNNLayer}
\end{figure}
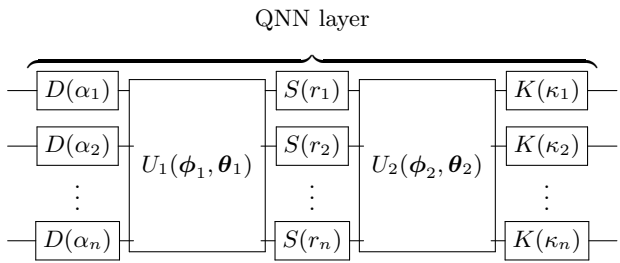

As suggested in \cite{Killoran2019}, we construct the quantum circuit shown in Fig.~\ref{fig:QNNLayer} as analogue of a single artificial neural network layer, where $\vect{\alpha}, \vect{r}, \boldsymbol{\kappa} \in \mathbb{R}^n$, $\vect{\phi}_1, \vect{\phi}_2 \in [0,2\pi]^n$ and $ \vect{\theta}_1, \vect{\theta}_2 \in [0,2\pi]^{n(n-1)/2}$ are the parameters of the circuit. $n(n-1)/2$ represents the number of beamsplitters for an $n$-mode universal interferometer \cite{universalInterferometer}. These adjustable parameters play the role of the weights and biases in a classical ANN. (Note that the non-linearity is also parametrized here.)

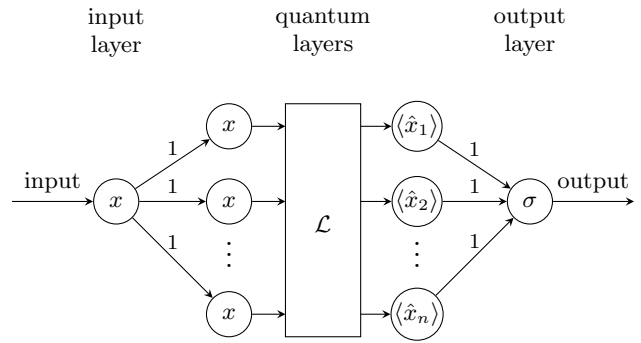
\begin{figure}[!ht]
\centering
\begin{tikzpicture}[>=stealth, neuron/.style={draw, circle, minimum size=0.6cm, inner sep=0}]
\node         (in)  at (-1.5, 0) {};
\node[neuron] (x)   at (0, 0)    {$x$};
\node[neuron] (xa)  at (1.5,  1) {$x$};
\node[neuron] (xb)  at (1.5,  0) {$x$};
\node (xc)  at (1.5,  -0.6) {\large \vdots};
\node[neuron] (xn)  at (1.5, -1.5) {$x$};
\draw (2.25, -1.8) rectangle (3.25, 1.3);
\node         (L)   at (2.75, -0.3) {$\mathcal{L}$};
\node[neuron] (ea)  at (4,  1)   {$\braket{\hat{x}_1}$};
\node[neuron] (eb)  at (4,  0)   {$\braket{\hat{x}_2}$};
\node (ec)  at (4,  -0.6) {\large \vdots};
\node[neuron] (en)  at (4, -1.5)   {$\braket{\hat{x}_n}$};
\node[neuron] (s)   at (5.5, 0)  {$\sigma$};
\node         (out) at (7, 0)    {};
\draw[->] (in) -- node[above] {input} (x);
\draw[->] (x)  -- node[above] {\footnotesize $1$} (xa);
\draw[->] (x)  -- node[above] {\footnotesize $1$} (xb);
\draw[->] (x)  -- node[above] {\footnotesize $1$} (xn);
\draw[->] (xa) -- (2.25,  1);
\draw[->] (xb) -- (2.25,  0);
\draw[->] (xn) -- (2.25, -1.5);
\draw[->] (3.25,  1) -- (ea);
\draw[->] (3.25,  0) -- (eb);
\draw[->] (3.25, -1.5) -- (en);
\draw[->] (ea) -- node[above] {\footnotesize $1$} (s);
\draw[->] (eb) -- node[above] {\footnotesize $1$} (s);
\draw[->] (en) -- node[above] {\footnotesize $1$} (s);
\draw[->] (s)  -- node[above] {output} (out);
\node[align=center] at (0,    2.25) {input \\ layer};
\node[align=center] at (2.75, 2.25) {quantum \\ layers};
\node[align=center] at (5.5,  2.25) {output \\ layer};
\end{tikzpicture}
\caption{Hybrid variational circuit for solving an ordinary differential equation. The input is broadcast to each mode before entering the QNN layer and is post-processed by a summation of the expectation values and (optionally) a classical activation function.}
\label{fig:hybrid_network}
\end{figure}

Inspired by the classical neural network architecture described in the previous section, we now embed such layers into a hybrid classical-quantum network, see Fig.~\ref{fig:hybrid_network}. The classical input $x$ is first broadcast and then embedded into each qumode using displacement gates. The output of the quantum layers is converted back to classical values by measurements of the position operator $\hat{x}$, which can be performed, e.g., using homodyne measurement \cite{leonhardt}. Expectation values can be estimated by repeated sampling. We emphasize that all adjustable parameters are contained in the quantum layers. For our simulations, the network consists of two qumodes and either one or two quantum layers. We sum the measurement averages of the qumodes and (optionally) apply a classical activation function to form the final output of the network.

For the present work, we perform simulations of such circuits and parameter optimization on a classical computer (see the following section). However, note that all steps could in principle also run on a physical CV quantum computer: using standard optical equipment to implement necessary gates, such circuits can be optimized by utilizing the parameter shift rule \cite{Li2017, Mitarai2018, Schuld}. This approach allows (under certain conditions) to obtain the gradient with respect to a parametrized gate by evaluating the circuit at two ``shifted'' values of the parameter, reminiscent of a difference quotient approximation but analytically exact (at least for \emph{Gaussian} CV circuits). Since the input $x$ is encoded via displacement gates, the parameter shift rule could be used for computing $y'(x)$ as well, and the second derivative in Eq.~\eqref{eq:partialDerivativeBackpropagation} by applying the rule twice. As caveat, we remark that this rule needs to be adapted and remains an approximation for circuits with non-Gaussian gates \cite{Schuld}. Moreover, these evaluations require many measurement samples to approximate expectation values, and have to be performed for each parameter separately. This poses high demands on the speed and accuracy of such measurements.   

We estimate the minimum (wall clock) time $T_\text{step}$ to run an encompassing optimization step in conjunction with the parameter shift rule on an actual quantum computer as follows. We denote the time of a pass through the circuit and subsequent measurement by $T_m$, the number of measurements for each expectation value by $M_\mu$, the number of parameters in the circuit by $M_p$, and the number of points within the domain $\Omega$ by $N$. $M_p$ depends on the number of layers and modes used in the particular architecture. Using universal interferometers with phaseless beamsplitters, we have $M_p = (n(n+4)) L$, where $L$ is the number of layers. To obtain the input gradient, a parameter shift rule has to be applied for each of the $n$ displacement embeddings according to the chain rule. We assume that the measurements in Fig.~\ref{fig:hybrid_network} can be performed in parallel. Taken together, one arrives at
\begin{equation}
\label{eq:computationalCost}
\begin{split}
T_\text{step} &\approx 4 n M_p L M_\mu N T_m\\
              &\approx 4 n^2(n+4) L M_\mu N T_m,
\end{split}
\end{equation}
where the factor 4 results from the nested composition of the parameter shift rule (gradient w.r.t.\ input $x$ and circuit parameters). 

As representative estimate, we insert $n=2$ for the number of modes, $L = 1$ for the number of layers, $N = 20$ for the number of points and $M_\mu = 100$ in order to get a reasonable expectation value into Eq.~\eqref{eq:computationalCost}, which gives $T_\text{step} \approx \num[round-precision=2,round-mode=figures]{192000}T_m$.

Using the numerical experiments in Sect.~\ref{sec:results} as guideline, the number of steps required for convergence is around $400$, so the total computational time would be $T_c\approx \num[round-precision=2,round-mode=figures]{76800000}T_m$.

\section{Simulation results}\label{sec:results}

We use the PennyLane software framework \cite{pennylane} with the Strawberry Fields \cite{Killoran2019strawberryfields, Bromley2020} \emph{Fock} backend by Xanadu for the simulations, with the occupancy cutoff set to $10$. We have chosen this cutoff dimension as large as possible while still keeping computational time low to increase the faithfulness of the simulation compared to a physical circuit. To keep the computational overhead low, two mode versions of Fig.\ref{fig:hybrid_network} with no classical activation function were used in all simulations. Computing the gradients of the network with respect to its parameters and input as well as circuit optimization is performed using the PyTorch interface and its automatic differentiation feature, together with PennyLane's implementation of the parameter shift rule. The ``training''  hyperparameters include learning rate, the specific optimizer used, the number of modes, the number of QNN layers and classical pre- and post-processing. The real number parameters were initialized by drawing from a normal distribution with mean 0 and standard deviation of 0.1 while the angle parameters were initialized from a uniform distribution with interval $[0,2\pi]$ using the \verb|cvqnn_layers_all| function from PennyLane. In order to  make the optimization deterministic, the seed argument was set to 0. 

\subsection{Linear differential equation}

We first apply the described approach to the IVP
\begin{equation}
y'(x) = - 2 x y(x), \quad y(0) = 1,
\label{eq:gaussianDiffEq}
\end{equation}
on the interval $x \in [-1, 1]$. As reference, the analytical solution can be found via separation of variables as
\begin{equation}
y(x) = e^{-x^2}.
\label{eq:gaussianDiffEqSolution}
\end{equation}

We tried several hyperparameters for best possible convergence, and achieved the best results with a learning rate of $0.02$ and the Adam optimizer.

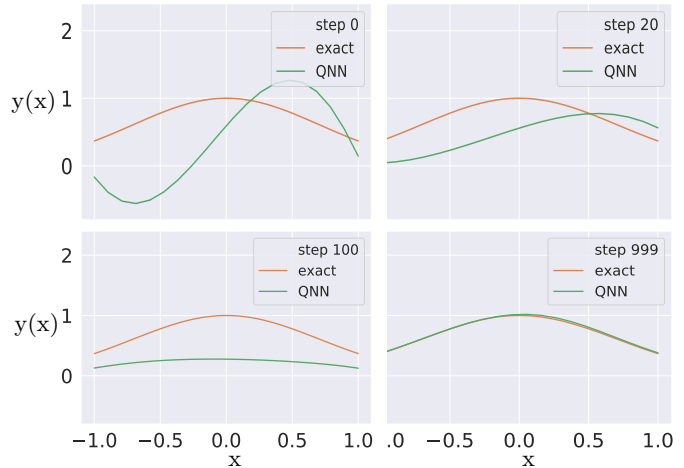
\begin{figure}[!ht]
  \centering
  \subfloat{
  \begin{tikzpicture}
  \node[inner sep=0pt] at (0,0)
  {\includegraphics[width=0.5\columnwidth,trim={0 1cm 0 0cm},clip,height=3.1cm]{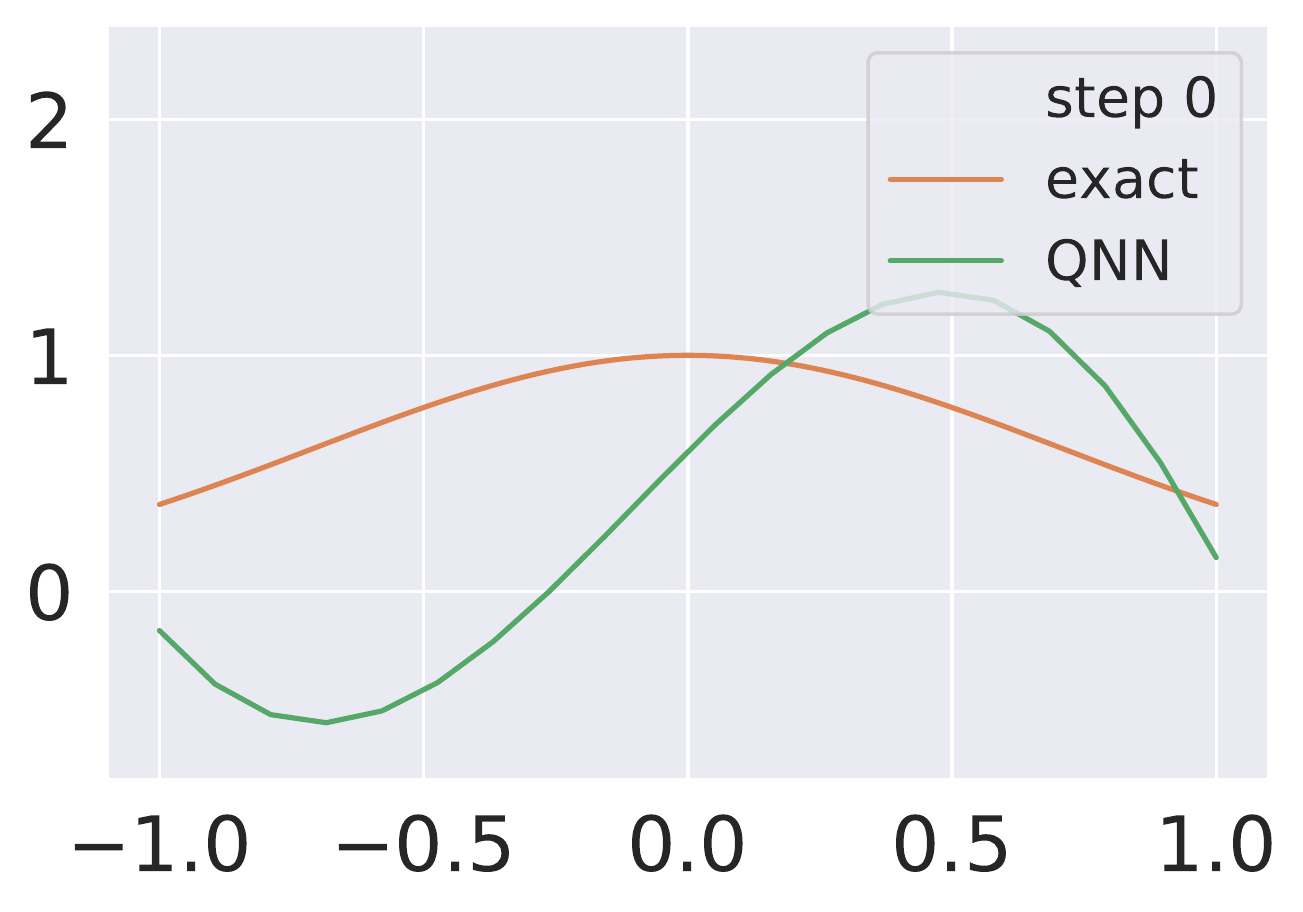}\label{fig:linearODEQNNConvergenceA}};
  \end{tikzpicture}}%
  \subfloat{
  \begin{tikzpicture}
  \node[inner sep=0pt] at (0,0)
  {\includegraphics[width=0.45\columnwidth,trim={1.9cm 1cm 0 0cm},clip,height=3.1cm]{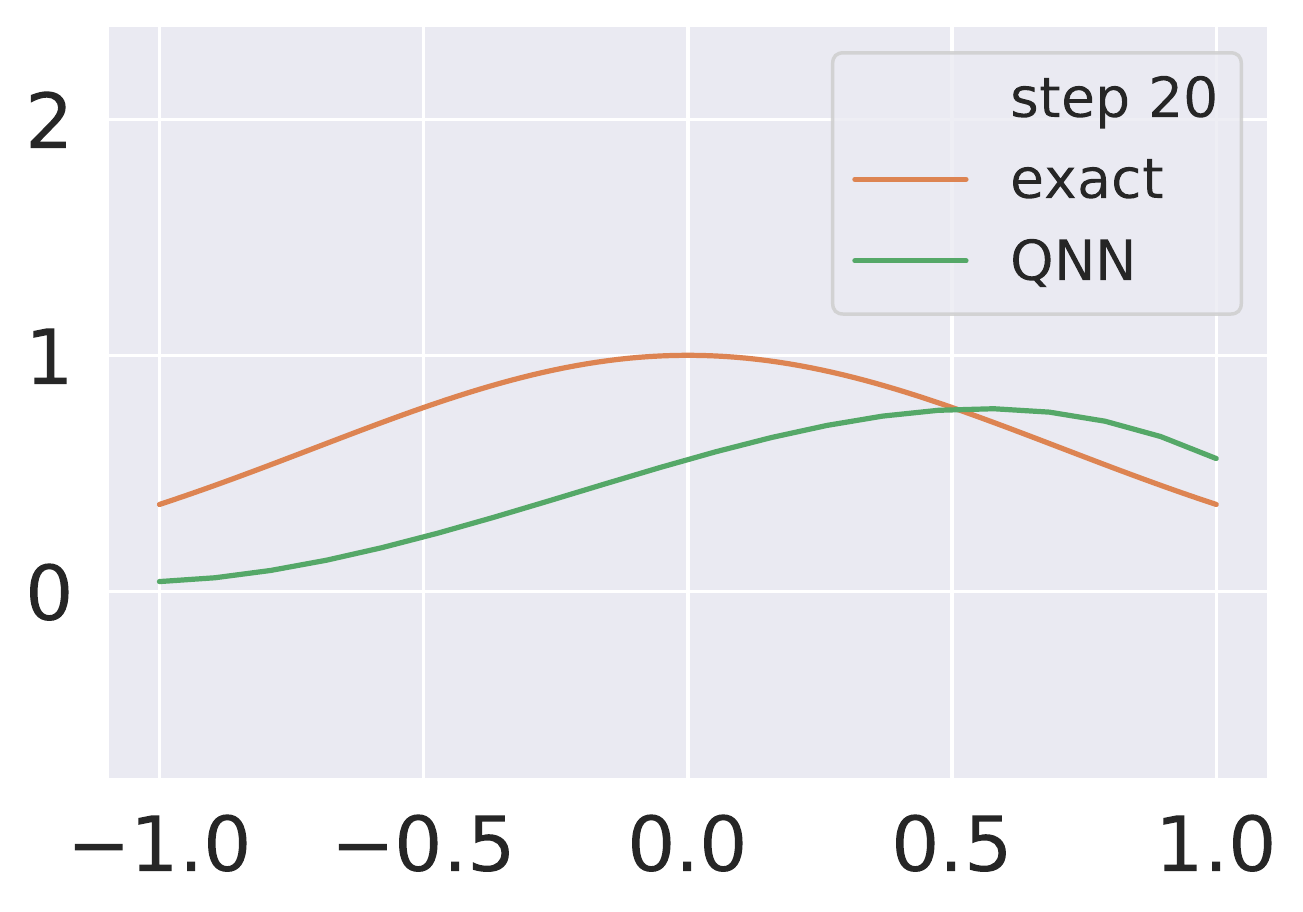}\label{fig:linearODEQNNConvergenceB}};
  \end{tikzpicture}}
  \put (-253,46) {y(x)}\\
  \vspace{-0.45cm}
  \subfloat{
  \begin{tikzpicture}
  \node[inner sep=0pt] at (0,0)
  {\includegraphics[width=0.5\columnwidth,trim={0 0 0 0cm},clip,height=3.1cm]{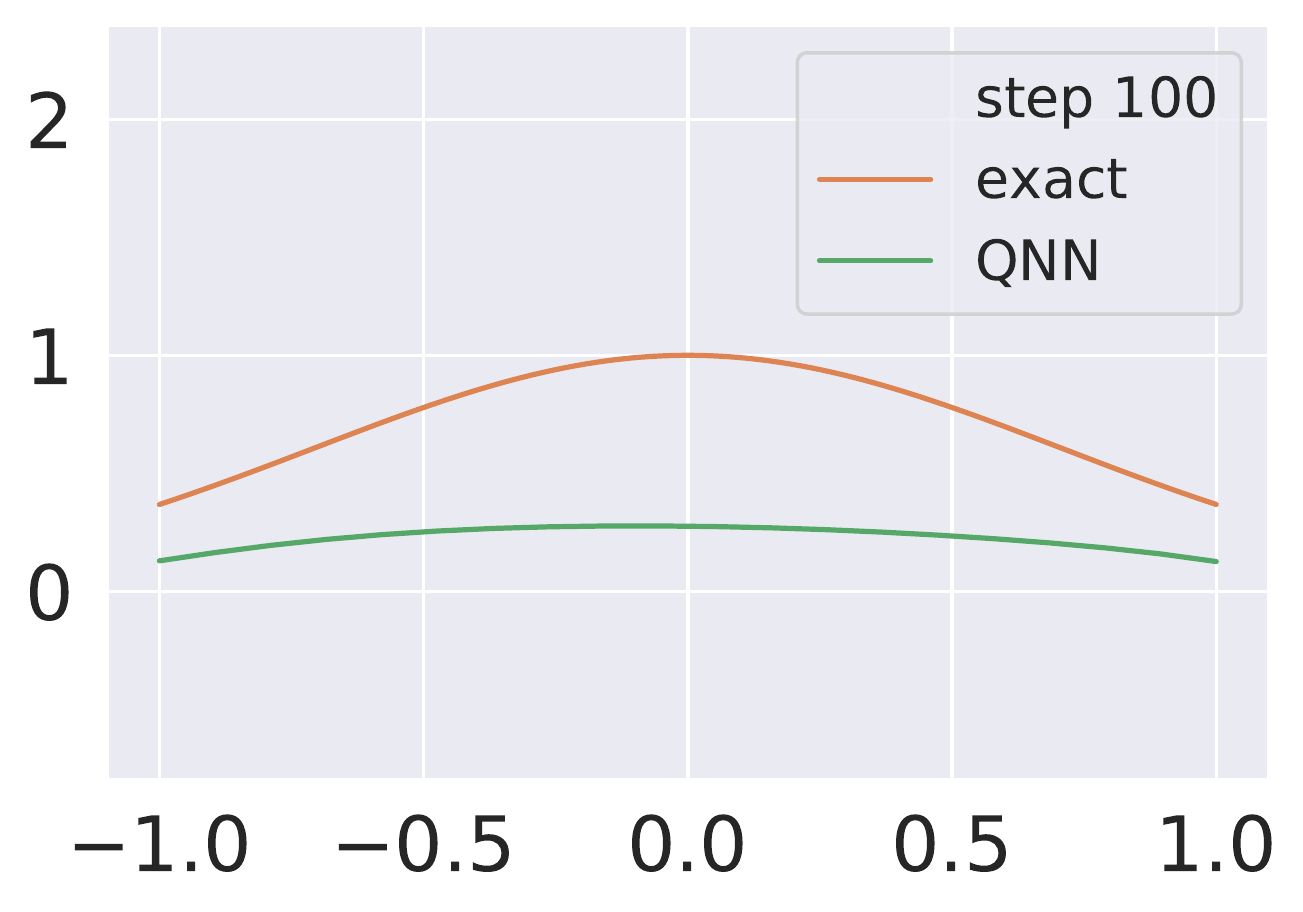}\label{fig:linearODEQNNConvergenceC}};
  \end{tikzpicture}}%
  \subfloat{
  \begin{tikzpicture}
  \node[inner sep=0pt] at (0,0)
  {\includegraphics[width=0.45\columnwidth,trim={1.9cm 0 0 0cm},clip,height=3.1cm]{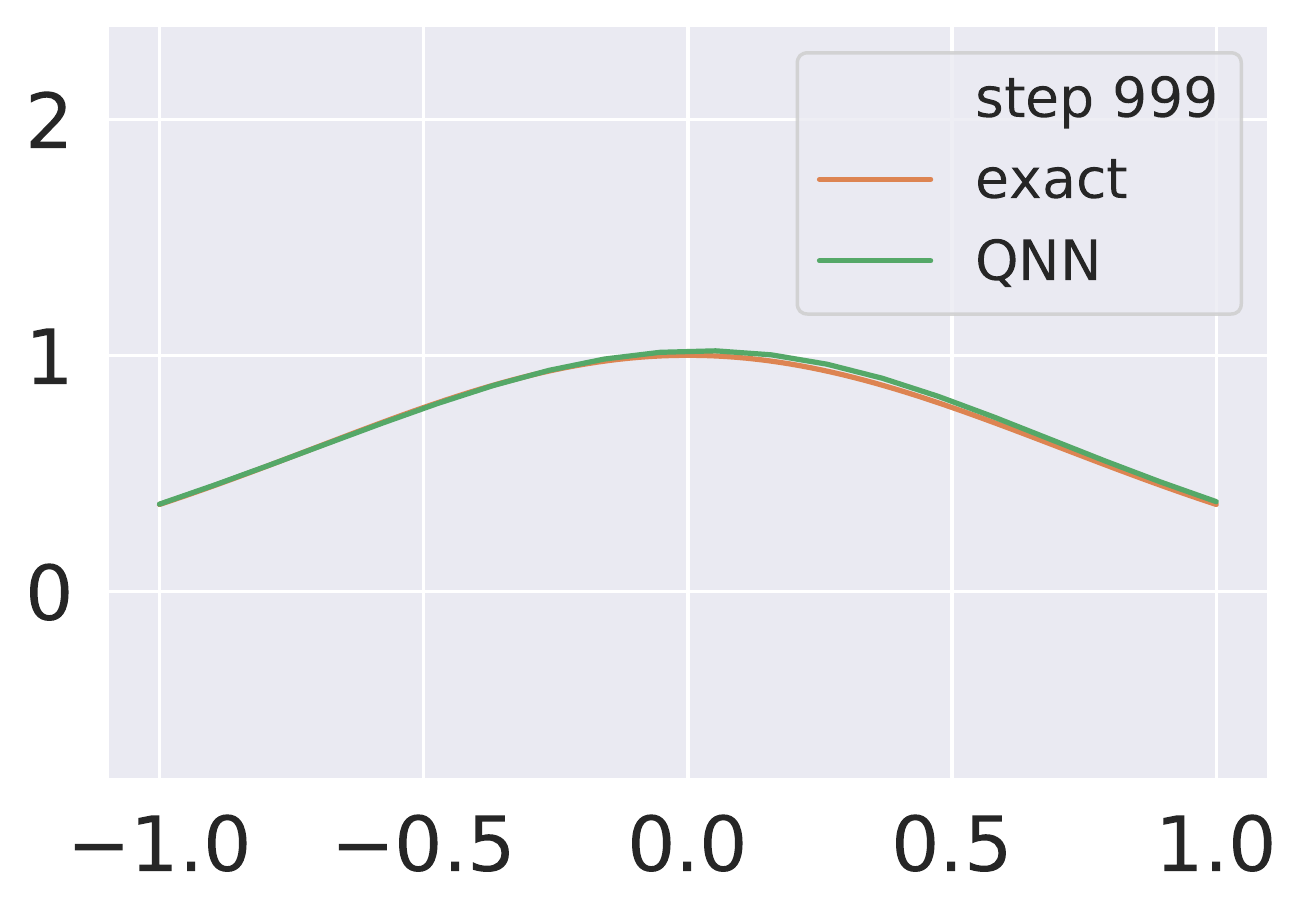}\label{fig:linearODEQNNConvergenceD}};
  \end{tikzpicture}}
\put (-255,48) { y(x)}
\put (-174,-3) { x}
\put (-63,-3) { x}
\caption{Approximate solution of the linear ODE \eqref{eq:gaussianDiffEq} via a 2 mode, 2 layer QNN as visualized in Fig.~\ref{fig:hybrid_network}.}
\label{fig:GaussianODEQNN}
\end{figure}

\begin{figure}[!ht]
  \centering
  \includegraphics[width=0.65\columnwidth]{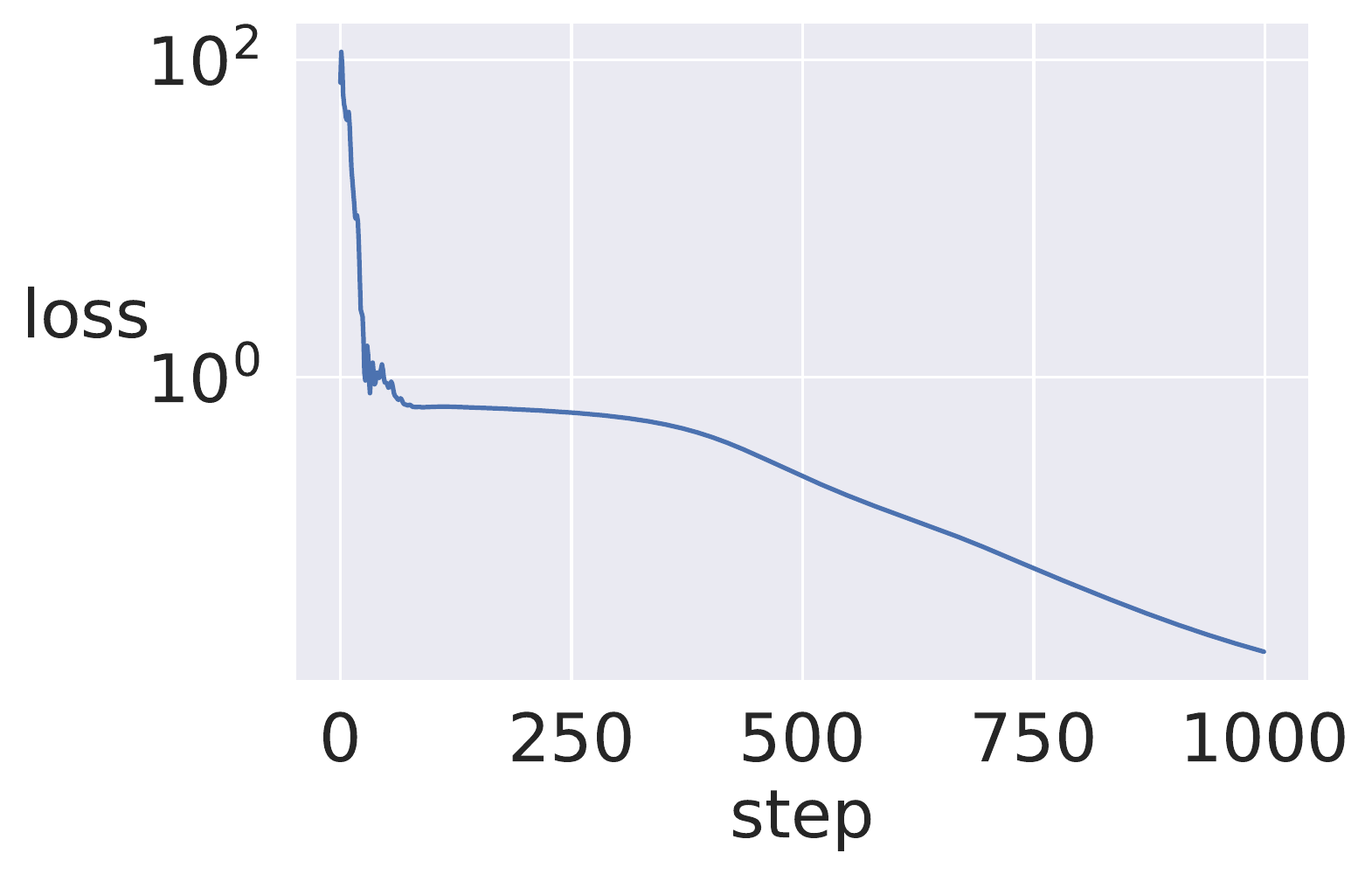}
\caption{Loss in Eq.~\eqref{eq:ODELossFunction} corresponding to the optimization shown in Fig.~\ref{fig:GaussianODEQNN}.}
\label{fig:linearODELoss}
\end{figure}

Several steps in the optimization can be seen in Fig.~\ref{fig:GaussianODEQNN}, and the resulting loss curve in Fig.~\ref{fig:linearODELoss}. A possible explanation for the optimization pattern are the two terms in the cost function \eqref{eq:ODELossFunction}, corresponding to the initial value constraint and satisfying $\frac{\ud y}{\ud x} = - 2 x y$. At step 100 the latter seems to be approximately satisfied, and the main contribution to the loss stems from the initial value constraint. In step 999, the initial value has been satisfied as well.

\subsection{Nonlinear differential equation}

To test the network on a \emph{nonlinear} ODE, we solve the IVP
\begin{equation}
y'(x) = x^2 + y(x)^2 - 1\, , \quad y(0) = 0.
\label{eq:nonLinearODE}
\end{equation}
This is a so-called Riccati equation \cite{Riccati, RiccatiSolution}. We use the \verb|odeint| function from the SciPy library \cite{scipy} to obtain a numerically exact reference solution. 

\begin{figure}[!ht]
  \centering
  \subfloat{
  \begin{tikzpicture}
  \node[inner sep=0pt] at (0,0)
  {\includegraphics[width=0.5\columnwidth,trim={0 1cm 0 0cm},clip,height=3.1cm]{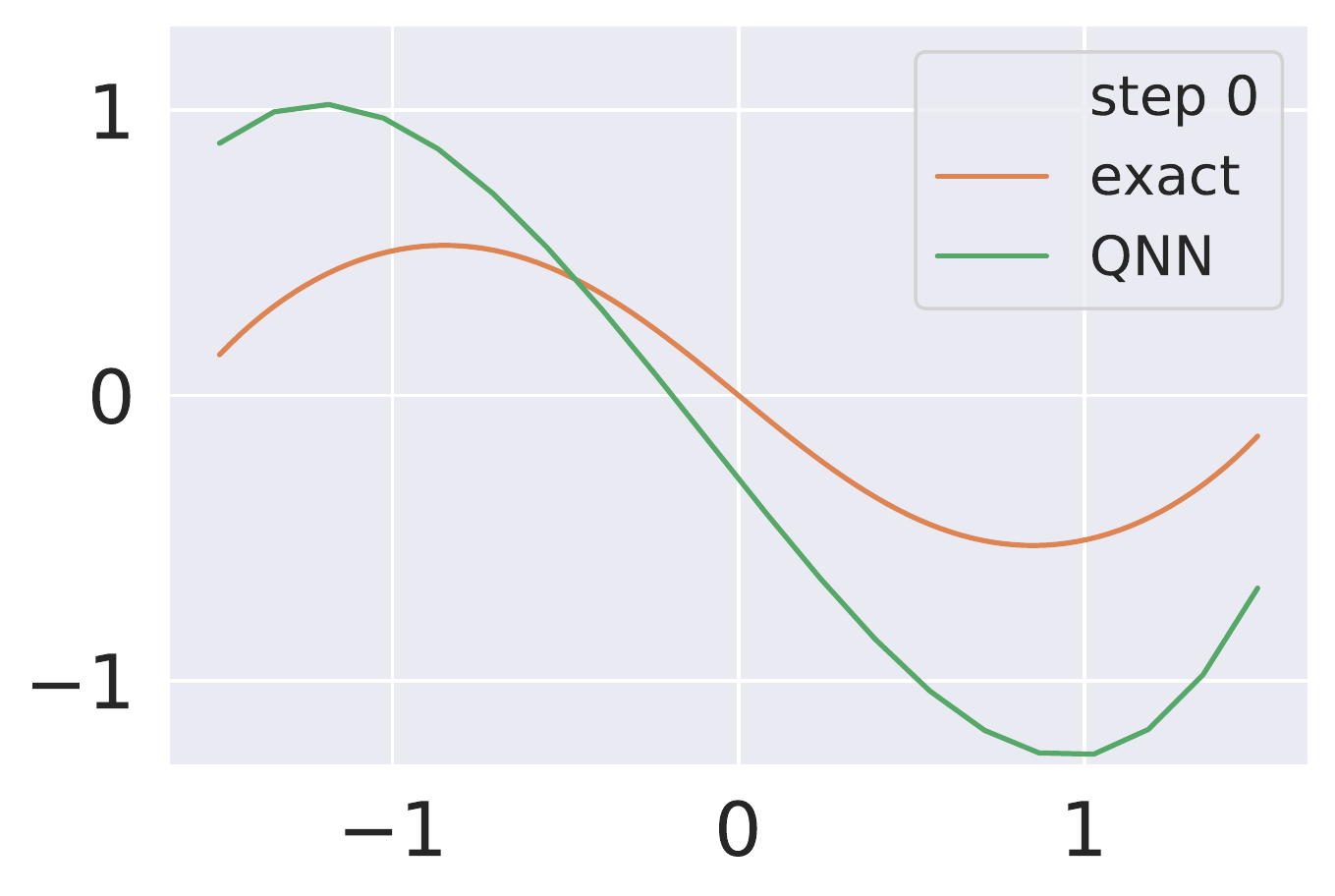}\label{fig:nonLinearODEQNNConvergenceA}};
  \end{tikzpicture}}%
  \subfloat{
  \begin{tikzpicture}
  \node[inner sep=0pt] at (0,0)
  {\includegraphics[width=0.45\columnwidth,trim={1.9cm 1cm 0 0cm},clip,height=3.1cm]{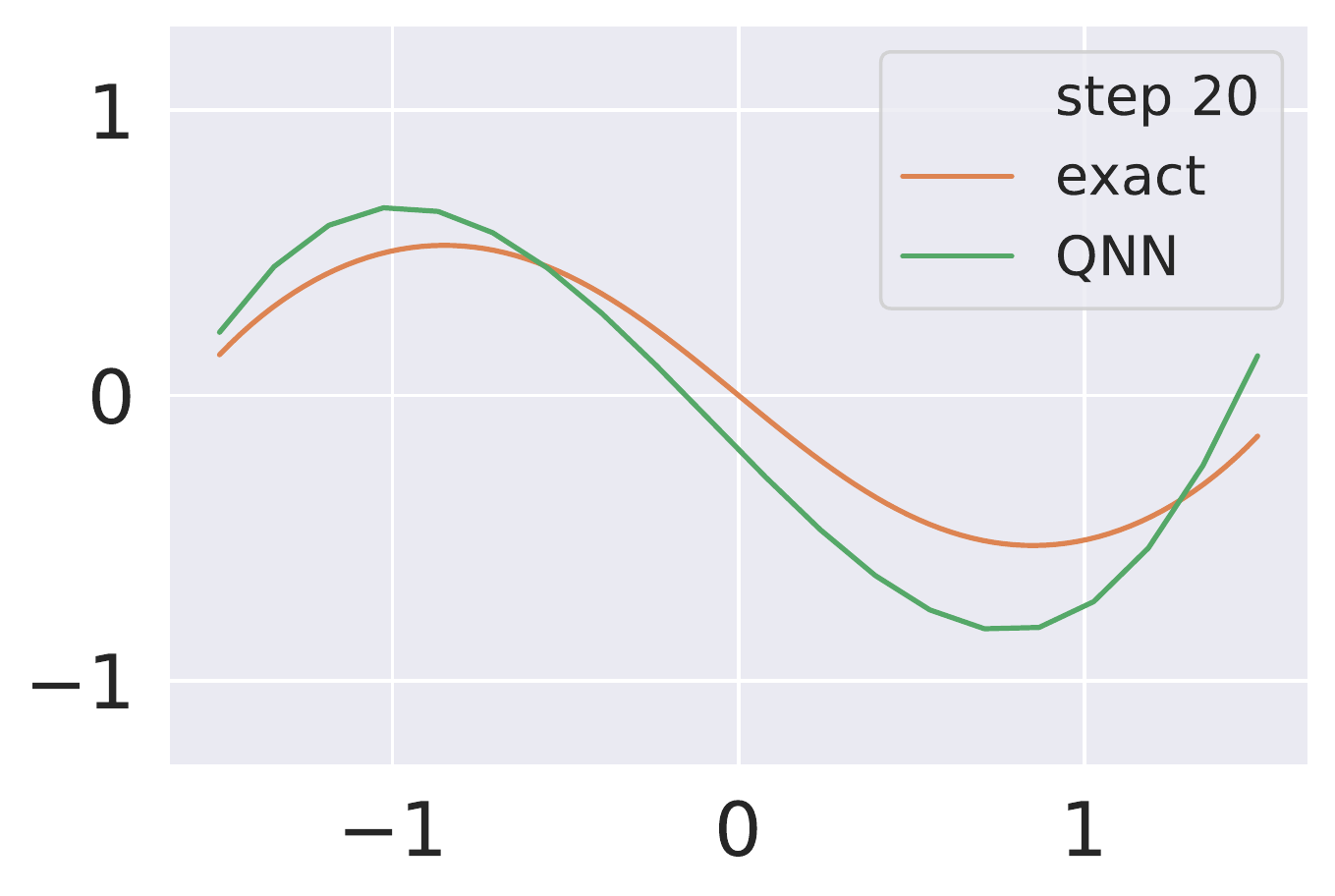}\label{fig:nonLinearODEQNNConvergenceB}};
  \end{tikzpicture}}
  \put (-247,46) {y(x)}\\
  \vspace{-0.45cm}
  \subfloat{
  \begin{tikzpicture}
  \node[inner sep=0pt] at (0,0)
  {\includegraphics[width=0.5\columnwidth,trim={0 0 0 0cm},clip,height=3.1cm]{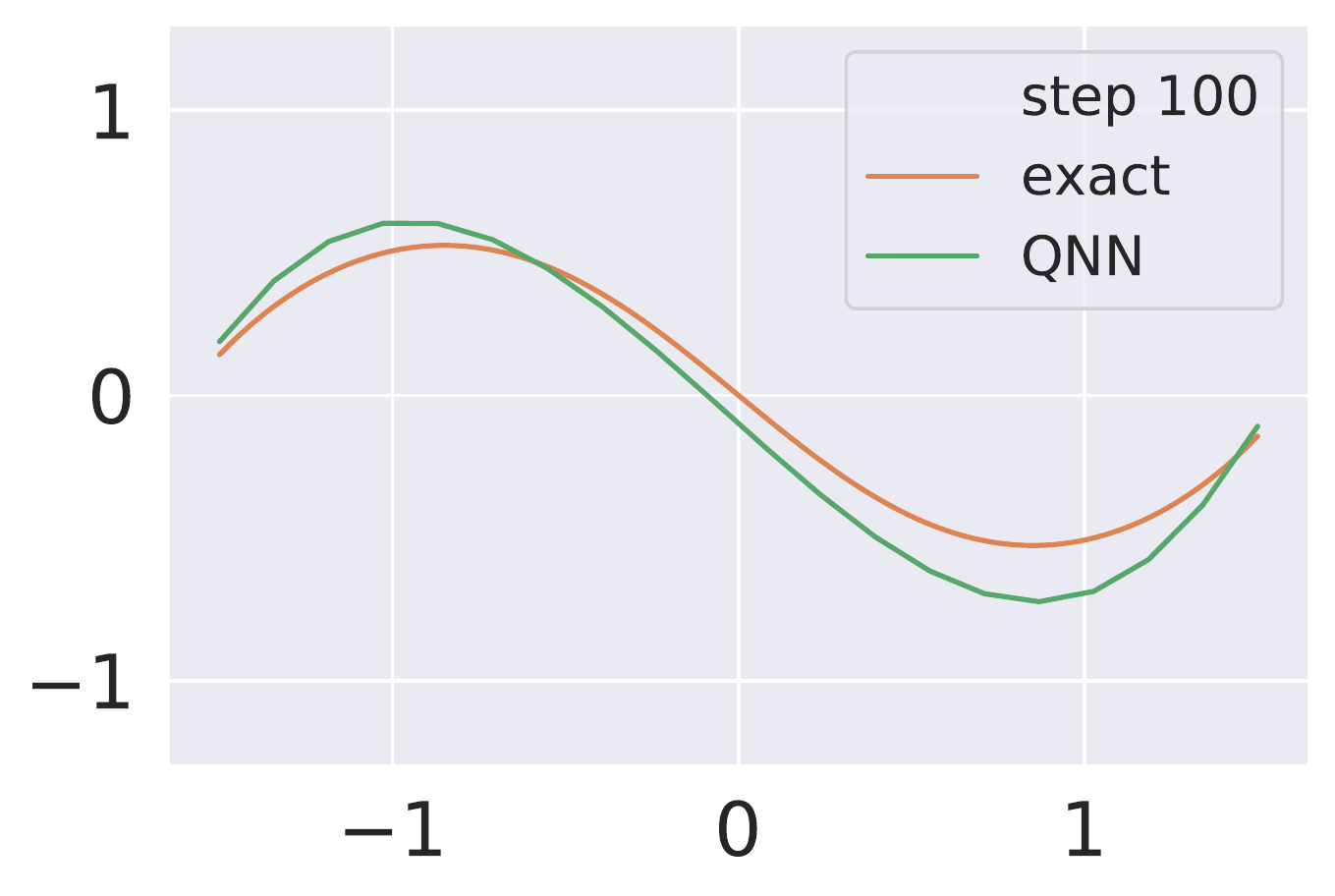}\label{fig:nonLinearODEQNNConvergenceC}};
  \end{tikzpicture}}%
  \subfloat{
  \begin{tikzpicture}
  \node[inner sep=0pt] at (0,0)
  {\includegraphics[width=0.45\columnwidth,trim={1.9cm 0 0 0cm},clip,height=3.1cm]{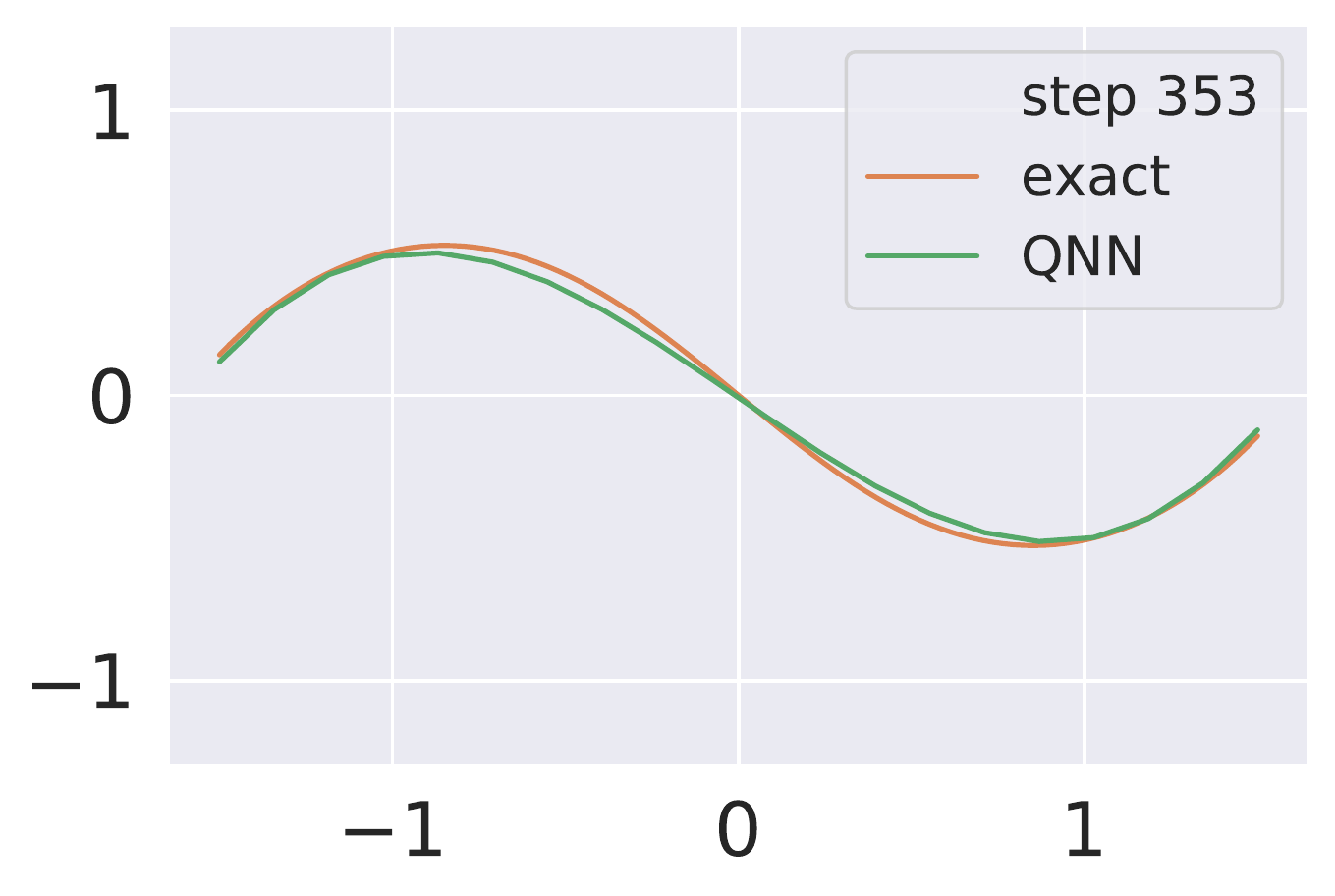}\label{fig:nonLinearODEQNNConvergenceD}};
  \end{tikzpicture}}
\put (-250,48) { y(x)}
\put (-174,-3) { x}
\put (-63,-3) { x}
\caption{Approximate solution of the ODE \eqref{eq:nonLinearODE} using a two-mode, single layer version of the variational circuit in Fig.~\ref{fig:hybrid_network}. }
\label{fig:nonLinearODEQNNConvergence}
\end{figure}
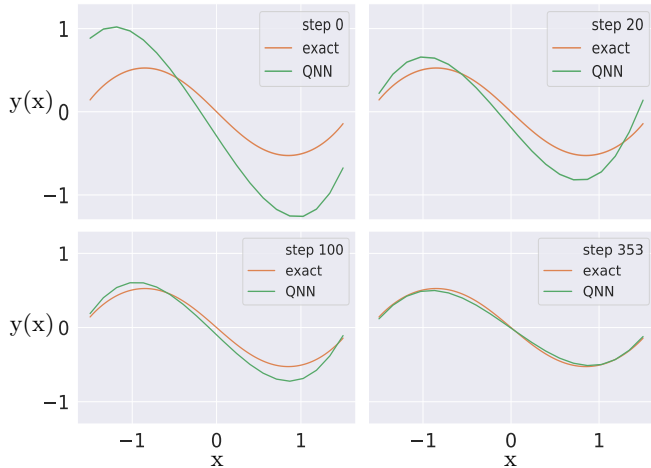

\begin{figure}[!ht]
  \centering
  \includegraphics[width=0.65\columnwidth]{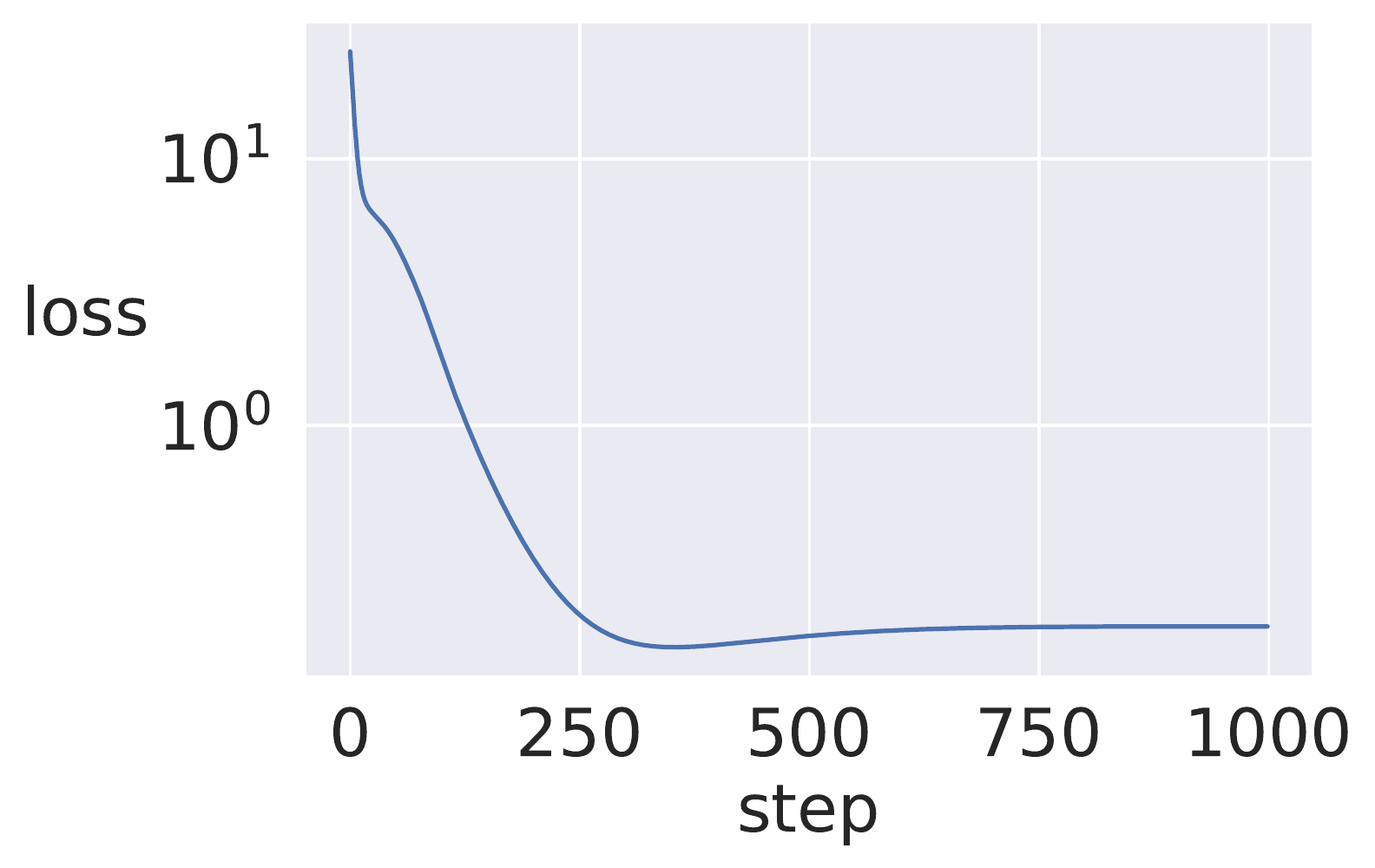}
\caption{Loss corresponding to the optimization in Fig.~\ref{fig:nonLinearODEQNNConvergence}.}
\label{fig:nonLinearODELoss}
\end{figure}

The result when using the Adam optimization strategy and a learning rate of $0.002$ can be seen in Fig.~\ref{fig:nonLinearODEQNNConvergence}, showing a good final approximation. Fig.~\ref{fig:nonLinearODELoss} visualizes the corresponding loss function. It seems to converge consistently until step 353, but then starts to increase slightly. Based on our numerical experiments, we have come to the conclusion that this is likely an artifact of the Adam optimizer close to the minimum loss \cite{AdamReason} in combination with the finite Fock occupancy cut-off, or an imbalance between the initial value and derivative terms in the cost function. To ameliorate the difficulty of the Adam optimizer, we added a constant to the denominator of the update rule, as suggested in \cite{AdamReason}, but this did not resolve the issue entirely.

\subsection{Stiff linear ODE}

Finally, we test our ansatz on the stiff ODE:
\begin{equation}
y'(x) = -2 y(x), \quad y(0) = \tfrac{1}{2}
\label{eq:stiffODE}
\end{equation}
which has the analytical solution
\begin{equation}
y(x) = \tfrac{1}{2}e^{-2x}.
\label{eq:stiffODESolution}
\end{equation}

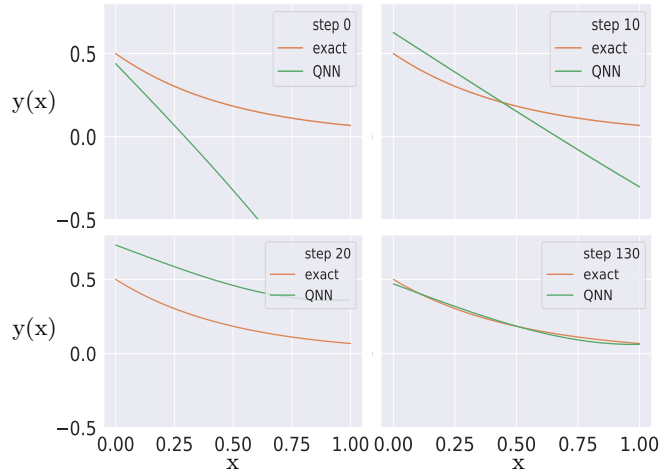
\begin{figure}[!ht]
  \centering
  \subfloat{
  \begin{tikzpicture}
  \node[inner sep=0pt] at (0,0)
  {\includegraphics[width=0.5\columnwidth,trim={0 1cm 0 0cm},clip,height=3.1cm]{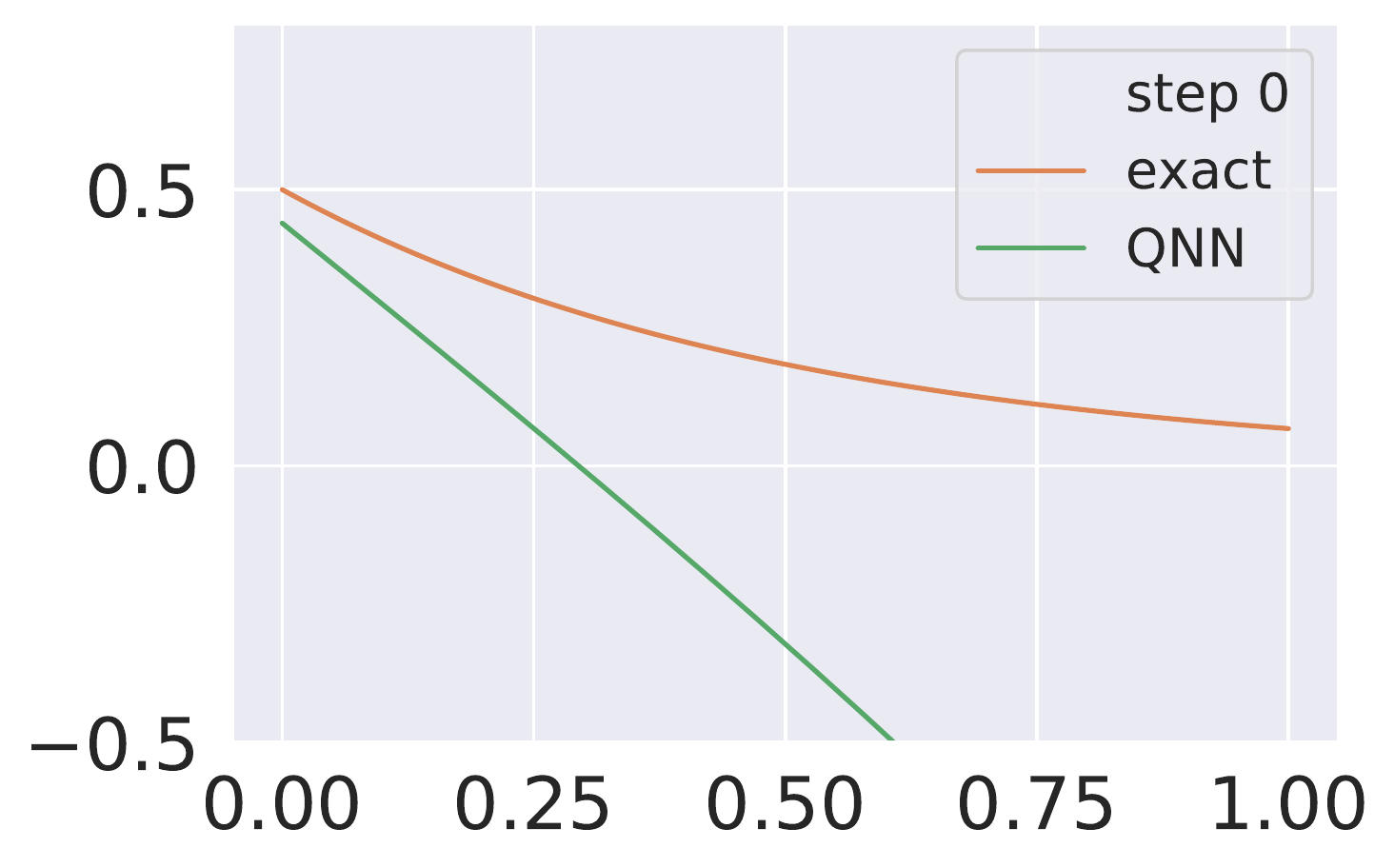}\label{fig:stiffODEQNNConvergenceA}};
  \end{tikzpicture}}%
  \hspace{-0.25cm}
  \subfloat{
  \begin{tikzpicture}
  \node[inner sep=0pt] at (0,0)
  {\includegraphics[width=0.45\columnwidth,trim={2.1cm 1cm 0 0cm},clip,height=3.1cm]{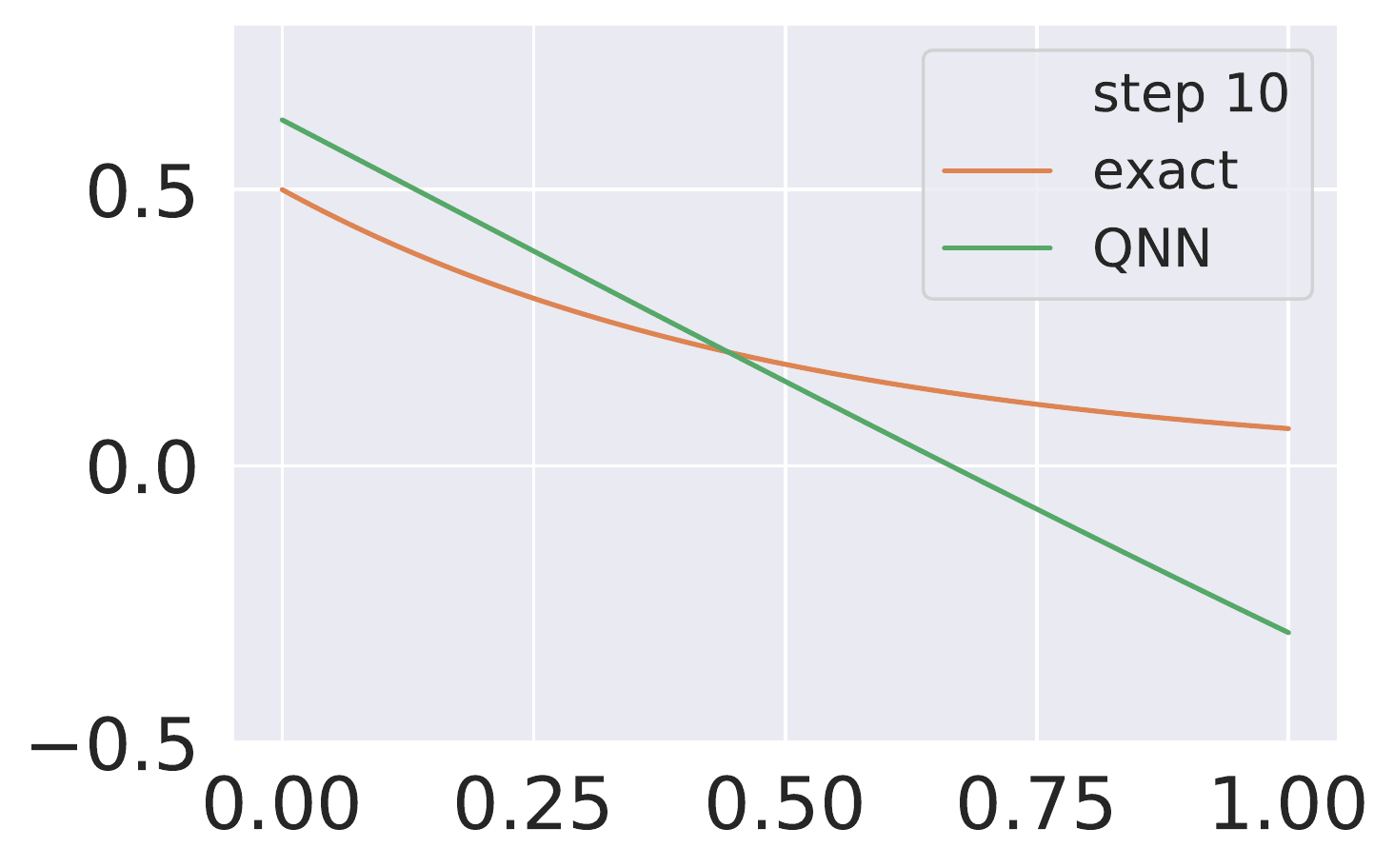}\label{fig:stiffODEQNNConvergenceB}};
  \end{tikzpicture}}
  \put (-247,46) {y(x)}\\
  \vspace{-0.4cm}
  \subfloat{
  \begin{tikzpicture}
  \node[inner sep=0pt] at (0,0)
  {\includegraphics[width=0.5\columnwidth,trim={0 0 0 0cm},clip,height=3.1cm]{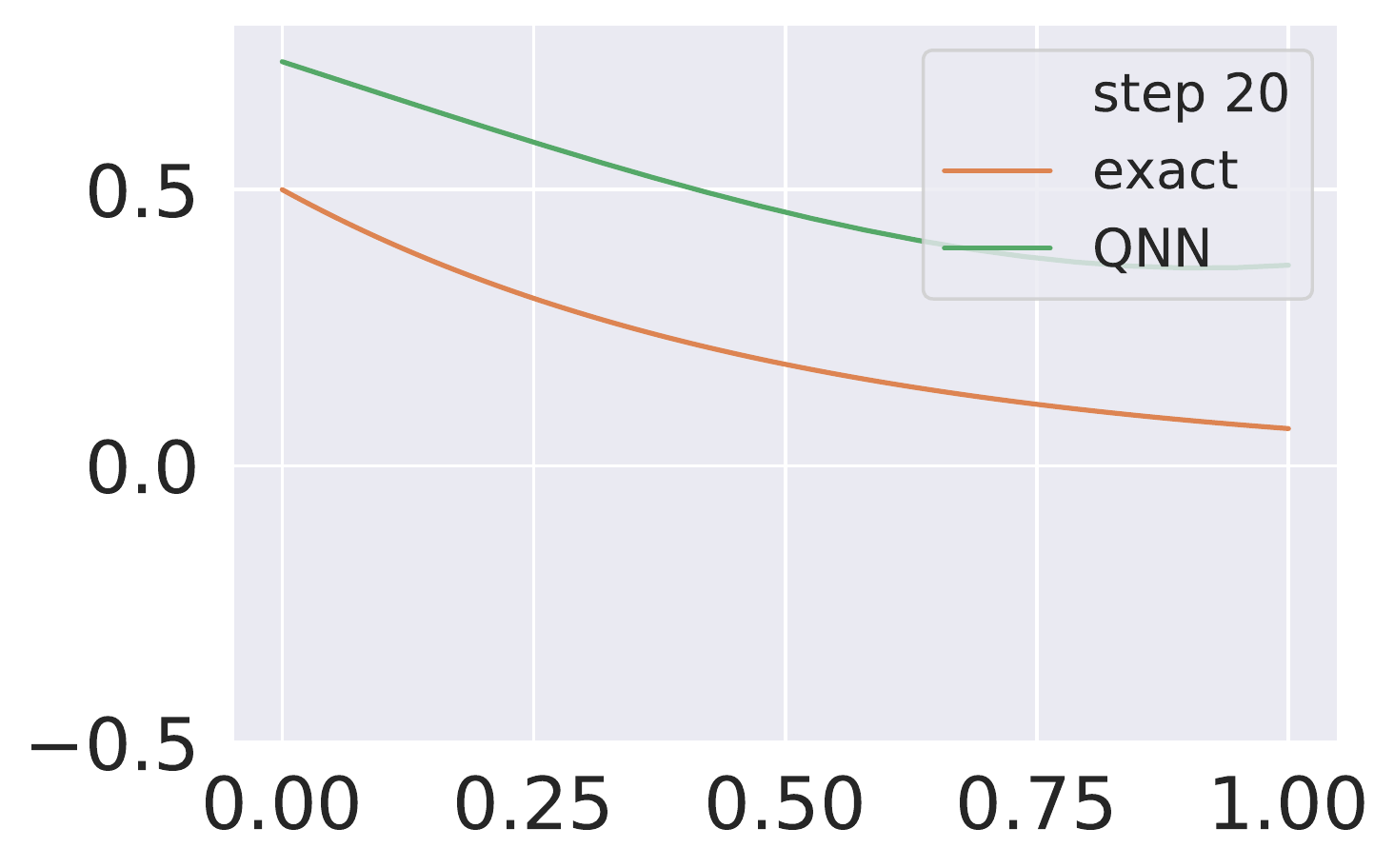}\label{fig:stiffODEQNNConvergenceC}};
  \end{tikzpicture}}%
  \hspace{-0.25cm}
  \subfloat{
  \begin{tikzpicture}
  \node[inner sep=0pt] at (0,0)
  {\includegraphics[width=0.45\columnwidth,trim={2.1cm 0 0 0cm},clip,height=3.1cm]{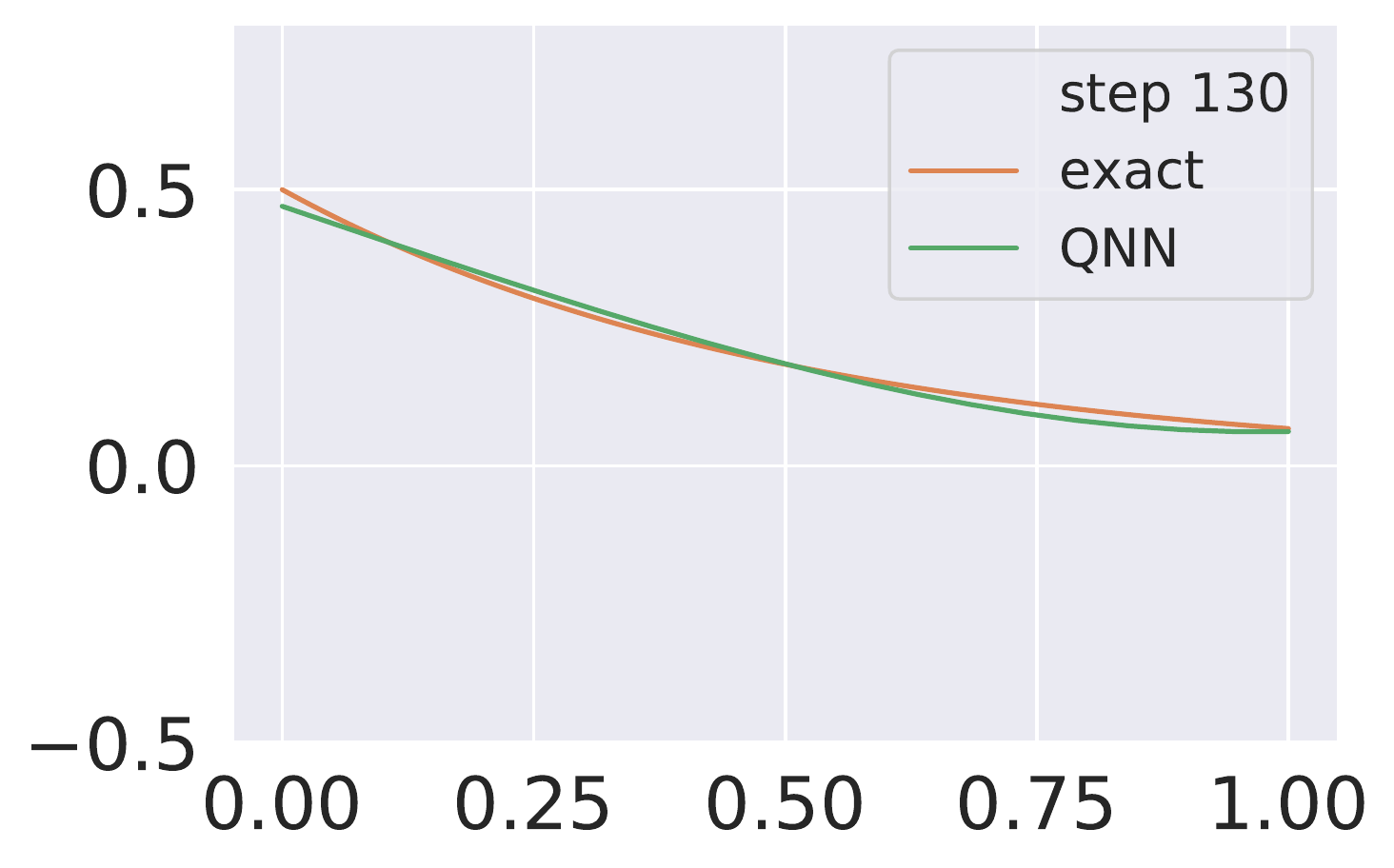}\label{fig:stiffODEQNNConvergenceD}};
  \end{tikzpicture}}
\put (-250,48) { y(x)}
\put (-169,-3) { x}
\put (-61,-3) { x}
\caption{Solution of the ODE \eqref{eq:stiffODE} using a 2 mode, 1 layer version of the variational circuit in Fig.~\ref{fig:hybrid_network}. }
\label{fig:stiffODEQNNConvergence}
\end{figure}

Here, we had success with a 2 mode, 1 layer architecture version of Fig.~\ref{fig:hybrid_network}. We again used the Adam optimizer, with a relatively small learning rate set to $0.005$. The result is visualized in Fig.~\ref{fig:stiffODEQNNConvergence}. A similar convergence pattern as for the nonlinear ODE was observed.  

\section{Conclusions}

The present work is based on simulation, and a natural question is how feasible and efficient the method could work on an actual CV quantum computer. A particular feature when considering differential equations is the appearance of second order derivatives in the cost function gradients with respect to the circuit parameters, see Eq.~\eqref{eq:partialDerivativeBackpropagation}. An alternative to the parameter shift rule discussed above could be a hybrid quantum-classical approach with a gradient-free parameter optimization. This might be more practicable in case the number of measurements required by the parameter shift rule becomes too demanding. A mixed strategy, where only the derivative w.r.t.\ the input $x$ is evaluated via the parameter shift rule, is also conceivable. A related interesting question in view of higher-order or partial differential equations concerns the accuracy and viability of repeated applications of the parameter shift rule to obtain higher-order derivatives.

Concerning function approximation in general, one aspect we leave for future research is whether the required variable ranges are feasible on a physical CV quantum computer, e.g., what are the minimum and maximum values of the $\hat{x}$ expectation values compatible with the experimental setup?

We consider \emph{analog} quantum computing applied to differential equations an interesting future direction, i.e., identifying and characterizing differential equations which are intrinsically solved by the time evolution of CV quantum systems.

\begin{acknowledgments}
We would like to thank Benjamin Zanger, Martin Schreiber and Martin Schulz for helpful discussions, and acknowledge support from the Munich Quantum Center and the TUM Institute for Advanced Study.
\end{acknowledgments}

\newpage

\appendix

\section{CV framework and gates}\label{sec:CV_gates}

CV quantum computing is typically realized by photons and optical instruments, and can in principle operate at room temperature. Cryogenic cooling is required for certain types of photon detectors, like superconducting nanowire single photon detectors (SNSPDs) \cite{SNSPD2014} or transition-edge sensors \cite{Irwin2020}.

As theoretical framework of optical quantum systems, one expands the electromagnetic potential of light in terms of its classical and quantum components \cite{leonhardt}
\begin{equation}
\mathbf{\hat{A}}(\mathbf{r},t) = \sum_k \left( \mathbf{A}_k(\mathbf{r},t) \hat{a}_k + \mathbf{A}_k^*(\mathbf{r},t) \hat{a}_k^\dagger\right),
\label{eq:modeExpansion}
\end{equation}
where $\mathbf{A}_k(\mathbf{r},t)$ are the classical complex wave functions, called modes, and the $\hat{a}_k$ are bosonic annihilation operators. These obey the characteristic bosonic commutation relations \cite{leonhardt}:
\begin{equation}
[\hat{a}_k,\hat{a}_{k'}^\dagger]=\delta_{kk'}, \quad [\hat{a}_k,\hat{a}_{k'}]=0.
\label{eq:boseRelations}
\end{equation}
The eigenstates of the annihilation operator are so-called coherent states \cite{leonhardt}, which is important for distinguishing gates.

One now defines position and momentum operators via
\begin{equation}
\hat{x}=\frac{1}{\sqrt{2}}\left(\hat{a}^\dagger+\hat{a}\right), \quad \hat{p}=\frac{i}{\sqrt{2}}\left(\hat{a}^\dagger-\hat{a}\right).
\label{eq:quadratures}
\end{equation}
They can be interpreted as real and imaginary parts of the annihilation operator, since
\begin{equation}
\hat{a} = \frac{1}{\sqrt{2}} \left(\hat{x}+i\hat{p}\right).
\label{eq:altitudeOfQuadratures}
\end{equation}
Eq.~\eqref{eq:boseRelations} implies $[\hat{x}, \hat{p}] = i$, i.e., they are canonically conjugate, as expected.

Next, we summarize the Gaussian CV gates used in this work: displacement, rotation, squeeze and beamsplitter gates \cite{OpticalQuantumInformationBook, Killoran2019}. These gates, when acting on a Gaussian (coherent)  state, return another Gaussian state. Although they are properly understood as working on coherent states, they have simple effects on the position and momentum operators. The displacement gate adds a constant. The (phase) rotation gate, as defined here, performs a clockwise rotation in phase space and the squeeze gate results in a scaling. The phaseless beamsplitter is a basic two-mode gate and results in an interference between the modes. The following equations summarize their respective effects on the quadrature operators:
\begin{align}
D(\alpha):\ 
\begin{pmatrix} \hat{x} \\ \hat{p} \end{pmatrix}
&\mapsto
\begin{pmatrix}
\hat{x}+\sqrt{2}\operatorname{Re}(\alpha)\\
\hat{p}+\sqrt{2}\operatorname{Im}(\alpha)
\end{pmatrix}, \quad \alpha \in \C,
\label{eq:displacementGate}\\
R(\phi):\ 
\begin{pmatrix} \hat{x} \\ \hat{p} \end{pmatrix}
&\mapsto
\begin{pmatrix}
\cos{\phi} & \sin{\phi}\\
-\sin{\phi} &\cos{\phi} 
\end{pmatrix}
\begin{pmatrix} \hat{x} \\ \hat{p} \end{pmatrix}, \ \phi \in [0, 2\pi],
\label{eq:phaseGate}\\
S(r):\ 
\begin{pmatrix} \hat{x} \\ \hat{p} \end{pmatrix}
&\mapsto
\begin{pmatrix}
e^{-r} & 0 \\
0 & e^{r} 
\end{pmatrix}
\begin{pmatrix} \hat{x} \\ \hat{p} \end{pmatrix}, \quad r \in \R
\label{eq:squeezeGate}\\
\mathrm{BS}(\theta):\ 
\begin{pmatrix} \hat{x}_1 \\ \hat{x}_2 \\ \hat{p}_1 \\ \hat{p}_2 \end{pmatrix}
&\mapsto
\begin{pmatrix}
\cos{\theta} & -\sin{\theta}&0&0\\
\sin{\theta} &\cos{\theta}&0&0\\
0&0&\cos{\theta} & -\sin{\theta}\\
0&0&\sin{\theta} &\cos{\theta}
\end{pmatrix}
\begin{pmatrix} \hat{x}_1 \\ \hat{x}_2 \\ \hat{p}_1 \\ \hat{p}_2 \end{pmatrix},
\label{eq:beamSplitGate}
\end{align}
where $\theta \in [0, 2\pi]$. As a non-linearity, the Kerr gate \cite{OpticalQuantumInformationBook} is used in this work, which has the operator representation
\begin{equation}
K(\kappa) = \e^{i\kappa\hat{n}^2},
\label{eq:kerrGate}
\end{equation}
with $\hat{n} = \hat{a}^{\dagger} \hat{a}$. Its effect on the position and momentum operators can be summarized by
\begin{equation}
K(\kappa):\ 
\begin{pmatrix} \hat{x} \\ \hat{p} \end{pmatrix}
\mapsto
\begin{pmatrix}
\hat{x}\cosh{\phi}-i\hat{p}\sinh{\phi}\\
\hat{p}\cosh{\phi}+i\hat{x}\sinh{\phi}
\end{pmatrix},
\label{eq:kerrGateEffectX}
\end{equation}
where $\phi(\hat{x},\hat{p}) = \kappa (\hat{x}\hat{p} - \hat{p}\hat{x} - i\hat{x}^2 - i\hat{p}^2)$. Notice that this transformation is non-linear in the operators, which is an important feature regarding the realization of artificial neural networks.

\bibliographystyle{unsrtmod}
\bibliography{literature}

\end{document}